# HEAT HYPERBOLIC DIFFUSION IN PLANCK GAS

M. Kozłowski*

and

J. Marciak – Kozłowska

Institute of Electron Technology, Al. Lotników 32/46

02 – 668 Warsaw, Poland

**Abstract**

In this paper we investigate the diffusion of the thermal pulse $e^{i\omega t}$ in Planck Gas. We show that the Fourier diffusion equation gives the speed of diffusion, *v > c* and breaks the causality of the thermal processes in Planck gas . For hyperbolic heat transport v<c and causality is valid

## 1 INTRODUCTION

In this paper we develop a description of heat transport in Planck gas We consider the Planck gas of the Planck particles i.e. particles with the masses equal to Planck mass~$10^{-5}$ g  Within the context of special relativity theory we investigate the Fourier and hyperbolic diffusion equations. We calculate the speed of thermal diffusion in the Fourier approximation and show that the thermal diffusion velocity exceeds the speed of light. We show that this result breaks the causality The same phenomena we describe within the framework of the hyperbolic thermal diffusion equation and show that in that case the speed of diffusion is always less than the speed of light.

## 2 MINKOWSKI SPACE-TIME

We may use the concept that the speed of light *in vacuo* is the upper limit of speed and also in which a signal can travel between two events to establish whether or not any two events could be connected. In the interest of simplicity we shall work with one space dimension $x_1 = x$ and the time dimension $x_o = ct$ of the Minkowski space-time. Let us consider events (1) and (2): their Minkowski interval *Δs* satisfies the relationship:

$$\Delta s^2 = c^2 \Delta t^2 - \Delta x^2 \qquad (1)$$

Without loss of generality we can take Event 1 to be at $x = 0$, $t = 0$. Then Event 2 can be only related to Event 1 if it is possible for a signal travelling at the speed of light, to connect them. If Event 2 is at (*Δx, cΔt*), its relationship to Event 1 depends on whether *Δs* > 0, = 0, or < 0.

We may summarise the three possibilities as follows:

Case A  *time-like* interval, $|\Delta x_A| < c\Delta t$, or $\Delta s^2 > 0$. Event 2 can be related to Event 1,

      Events 1 and 2 can be in causal relation.

Case B  *light-like* interval, $|\Delta x_B| = c\Delta t$, or $\Delta s^2 = 0$. Event 2 can only be related to
      Event 1 by a light signal.

Case C  *space-like* interval $|\Delta x_A| > c\Delta t$, or $\Delta s^2 < 0$. Event 2 cannot be related to

Event 1, for in that case *v > c*.

Now let us consider case C in more detail. At first sight, it seems that in case C we can find out the reference frame in which two Events $c^>$ and $c^<$ always fulfils the relationship $t_{c^>} - t_{c^<} > 0$. but this is not true. If we choose the inertial frame U' in which $t'_{c^>} - t'_{c^<} > 0$ and the reference frame U is moving with a speed *V* relative to U', with

$$V = c \frac{c(t'_{c^>} - t'_{c^<})}{x'_{c^<} - x'_{c^>}} \tag{2}$$

then for a speed *V<c*

$$\left| \frac{c(t'_{c^>} - t'_{c^<})}{x'_{c^<} - x'_{c^>}} \right| < 1 \tag{3}$$

Let us calculate $t_{c^>} - t_{c^<}$ in the reference frame U

$$t_{c^>} - t_{c^<} = \frac{1}{\sqrt{1 - \frac{V^2}{c^2}}} \left[ \frac{V}{c^2}(x'_{c^>} - x'_{c^<}) + (t'_{c^>} - t'_{c^<}) \right] =$$

$$\frac{1}{\sqrt{1 - \frac{V^2}{c^2}}} \left[ \frac{t'_{c^>} - t'_{c^<}}{x'_{c^>} - x'_{c^<}} (x'_{c^>} - x'_{c^<}) + (t'_{c^>} - t'_{c^<}) \right] = 0 \tag{4}$$

For a higher *V* we have $t_{c^>} - t_{c^<} < 0$. This implies that for space-like intervals the sign of $t_{c^>} - t_{c^<}$ depends on the speed *V*, i.e. the causality relation for space-like events is not valid.

## 3   FOURIER DIFFUSION EQUATION AND SPECIAL RELATIVITY

In paper [1] the speed of diffusion signals was calculated

$$v = \sqrt{2D\omega} \tag{5}$$

where

$$D = \frac{\hbar}{m} \tag{6}$$

and ω is the angular frequency of the pulses. Considering equations (5) and (6) one obtains

$$v = c\sqrt{2\frac{\hbar\omega}{mc^2}} \quad (7)$$

and $v \geq c$ for $\hbar\omega \geq mc^2$.

From equation (7) we conclude that for $\hbar\omega > mc^2$ the Fourier diffusion equation is in contradiction with the special relativity theory and thus breaks the causality in transport phenomena.

## 4  HYPERBOLIC DIFFUSION AND SPECIAL RELATIVITY

In monograph [2] the hyperbolic model of the thermal transport phenomena was formulated. It was shown that the description of the ultra-short thermal energy transport needs the hyperbolic diffusion equation (one dimensional transport)

$$\tau\frac{\partial^2 T}{\partial t^2} + \frac{\partial T}{\partial t} = D\frac{\partial^2 T}{\partial x^2} \quad (9)$$

In equation (9) $\tau = \frac{\hbar}{m\alpha^2 c^2}$ is the relaxation time, $m$= Planck mass $\alpha$ is the coupling constant and $c$ is the speed of light in vacuum, $T(x,t)$ is the temperature field and $D = \hbar/m$.

In paper [1] the speed of thermal propagation $v$ was calculated

$$v = \frac{2\hbar}{m}\sqrt{-\frac{m}{2\hbar}\tau\omega^2 + \frac{m\omega}{2\hbar}(1+\tau^2\omega^2)^{1/2}} \quad (10)$$

Considering that $\tau = \hbar/m\alpha^2 c^2$ equation (10) can be written as

$$v = \frac{2\hbar}{m}\sqrt{-\frac{m}{2\hbar}\frac{\hbar\omega^2}{mc^2\alpha^2} + \frac{m\omega}{2\hbar}(1+\frac{\hbar^2\omega^2}{m^2 c^4 \alpha^4})^{1/2}} \quad (11)$$

For

$$\frac{\hbar\omega}{mc^2\alpha^2} < 1, \quad \frac{\hbar\omega^2}{mc^2} < 1 \quad (12)$$

one obtains from equation (11)

$$v = \sqrt{\frac{2\hbar}{m}\omega} \quad (13)$$

Formally equation (13) is the same as equation (7) but considering the inequality (11) we obtain

$$v = \sqrt{\frac{2\hbar\omega}{m}} = \sqrt{2}\alpha c < c$$

(14)

and the causality is not broken.

For

$$\frac{\hbar\omega}{mc^2} > 1; \qquad \frac{\hbar\omega}{\alpha^2 mc^2} > 1 \qquad (15)$$

we get from equation (11)

$v = \alpha c, \qquad v < c$ (16)

Considering equations (14) and (16) we conclude that the hyperbolic diffusion equation (9) describes the thermal phenomena in accordance with special relativity theory and the causality is not broken